\documentstyle[twocolumn,preprint,aps,prl]{revtex} 
\tighten                                  
\begin{document}                          
 \draft                                   

\title{Temperature dependence of surface reconstructions
of Au on Pd(110)}

\author{J. A. Nieminen}

\address{Department of Physics Tampere University
of Technology  \\ P.O.Box 692 FIN - 33101 Tampere
Finland}

\date{March 8, 1995}
\maketitle
\begin{abstract}
Surface reconstructions of Au film on Pd(110) substrate are
studied using a local Einstein approximation to quasiharmonic
theory with the Sutton-Chen interatomic potential. Temperature
dependent surface free energies
for different coverages and surface structures are calculated.
Experimentally observed transformations from $(1\times1)$ to
$(1 \times 2)$ and $(1 \times 3)$ structures can be explained
in the framework of this model. Also conditions for
Stranski-Krastanov growth mode are found to comply with experiments.
The domain of validity of the model neglecting mixing entropy
is analyzed.
\end{abstract}

Reconstructions of $(110)$-surfaces of {\it d}-metals have
inspired numerous experimental and theoretical studies.
It has been confirmed by Low Energy Electron Diffraction (LEED)
\cite{Noo} and Scanning Tunneling Microscopy (STM) \cite{Gri,Bin}
that {\it 4d}-metals, ($Rh$, $Pd$ and $Ag$) favour unreconstructed
$(1\times 1)$ structures whilst {\it 5d}-metals ($Ir$, $Pt$ and $Au$)
exhibit $(1\times2)$ reconstructions spontaneously. Reconstructions
of pure metals are theoretically fairly well
understood either in terms of the relative contribution of
the {\it d}-electrons to binding \cite{HoB}
or in terms of the range of interatomic potentials \cite{Hak,Tod}.
However, conditions for surface reconstructions in heterogenous
adsorbate-substrate systems have not been explained
in simple terms, yet.

A heterogenous system which has received both
experimental and theoretical attention is $Au$ on $Ag(110).$
Rousset et. al. \cite{Rou} have revealed with STM
that at submonolayer coverages $Au$ atoms
tend to migrate into the substrate so that a complete
$Ag$ layer is formed on the top
of the $Au$ adsorbate. They have also shown that at coverages,
$\Theta$, beyond a monolayer $(1ML)$ of $Au$
interdiffusion no longer happens but $Au$
atoms tend to form clusters on the top of the $Ag$ layer on
the surface. Thus $Au$ grows on $Ag(110)$ in a layer+island
or Stranski-Krastanov (SK) growth mode. The first principles
calculations by Chan et al. \cite{Cha} confirm that
it is energetically favourable for $Au$ atoms to be covered by
one $Ag$ layer at submonolayer regime. This has been
reproduced in Surface
Embedded Atom Model (SEAM) based Molecular dynamics (MD)
simulations, which also show an onset of
SK growth mode for coverages beyond $\Theta = 1ML$ \cite{Haf}.

In this Letter the temperature and coverage
dependence of $Au/Pd(110)$ system depicted from LEED studies
of Schmitz et. al \cite{Smi} is studied theoretically.
The structural phase transformations show four main features:
{\bf (i)} at submonolayer
coverages only $(1\times 1)$ structures are observed, {\bf (ii)}
for $\Theta \approx 1.5ML$ there is a transition from
$(1 \times 1)$ structure to a mixture of $(1\times 2)$ and
$(1\times3)$ domains at $T\approx 310K$ and back to $(1\times1)$
at $T \approx 530K$, {\bf (iii)} For $\Theta \approx 2ML$ a sequence
of transitions $(1\times1) \buildrel {310K} \over  \longrightarrow
(1\times n) \buildrel {640K} \over \longrightarrow (1\times2)
\buildrel {770K} \over \longrightarrow (1\times1)$
is seen, where $(1\times n)$ denotes a mixture of $(1\times 2)$
and $(1\times 3)$ domains; {\bf (iv)} increasing coverage beyond
$\Theta=2ML$ leads to ever fainter LEED patterns suggesting
clustering of $Au$ atoms on the top of two $Au$ layers,
i.e. growth in SK mode. The related LEED studies
by Kaukasoina et al. \cite{Kau} mainly confirm the previous
observations apart from two exceptions: they were not able to
reproduce the $(1\times 2)$ pattern for $\Theta = 1.5ML$ and
for $\Theta = 2.0ML$ they found a continuous transition
from $(1\times 1)$ pattern via $(1\times 2)$ to a mixture
of $(1\times2)$ and $(1\times3)$ patterns and then back to
$(1\times 1)$ structure. In addition, for $\Theta=2.5ML$
they found a $(1\times2)$ reconstructed surface with a concentration
profile $(1,0.9,0.36,0.67,0.35)$, i.e., atomic concentration of $Au$
is $100 \%$ in the first layer, $90 \%$ in the second, etc.

A feasible theoretical method to incorporate vibrational motion
into surface free energies is the local Einstein approximation
to quasiharmonic theory as formulated by Le Sar et al. \cite{LeS}:
\begin{equation}
F_E = E_P + 3kT\sum_{i} \ln{[{\hbar |D_i|^{1/6} \over
\sqrt{M_i} kT}]}
\end{equation}
where $E_P$ is the potential energy and $|D_i|$ is the determinant
of the $3 \times 3$ matrix
$ D_{i\alpha \beta} = {\partial^2 E_P \over
\partial u^{i}_{\alpha} \partial u^{i}_{\beta}}.$
At surfaces, where the three-dimensional symmetry breaks down,
it is more advantageous to use the local Einstein approximation
with off-diagonal elements of $D_{i\alpha \beta}$ than a
Debye-type approximation \cite{Sut1},
which utilizes the trace of $D_{i\alpha \beta}$ and
neglects shear properties.
It has been shown in Ref. \cite{Hai} in a study
of grain boundaries in $Si$, that although
Eq. (1) may fail in the absolute values of free energies,
the excess free energies
between two different structures are remarkably good.
An optimal structure is obtained by allowing the system
relax to a local minimum of Eq. (1).
In this study a modification of Polak-Ribiere optimization method
as expressed in Ref. \cite{Num}, is used with
numerical approximants of the gradient of free energy with respect to
atomic positions.

It has been shown that the Sutton-Chen (SC)
potential predicts the reconstructions of (110)-surfaces
of {\it d}-metals correctly
except for $Ir$ \cite{Tod}. In addition, SC-potential
is easy to apply in Eq. (1), in practice, and there exist
simple but physically satisfactory rules to describe bonding
between atoms of different species.
SC-potential \cite{Sut} is written as follows
\begin{equation}
E_{SC} = \sum_i\epsilon_{i}\{ {1 \over 2} \sum_{j \ne i}
({a \over r_{ij}})^n - c_{i}
\sqrt{ \sum_{j\ne i} ({a \over r_{ij}})^m} \},
\end{equation}
where $(m,n)= (7,12)$ and $(8,10)$,
$\epsilon = 4.1790 \times 10^{-3} eV$
and $1.2793 \times 10^{-2} eV$, $a= 3.89 \AA$ and $ 4.08 \AA$,
and $c = 108.27$ and $ 34.408$ for $Pd$ and $Au$, respectively.
In heterogenous $AB$-systems the following rules
$m^{AB} = {1 \over 2}
(m^{A}+m^{B})$, $n^{AB} = {1 \over 2}(n^{A}+n^{B})$,
$a^{AB} = \sqrt{a^{A}a^{B}}$ and $\epsilon^{AB} =
\sqrt{\epsilon^{A}\epsilon^{B}}$ are found to give satisfactory
structural, energetic and elastic properties \cite{Raf}.

The calculations were carried out using a slab of 14 layers in
the [110]-direction, each layer consisting of 6 rows and 6 columns
($36 \times 14 = 504$ atoms altogether) with periodic boundary
conditions in the direction of both rows and colums.
The size of the calculational cell is
limited by the cut-off radius of the SC potential, $r_{c} = 2a$,
and the number of rows must be divisible by
$2$ and $3$ for $(1 \times n)$
reconstructions with $n$ either $2$ or $3$ and the slab must be
thick enough to model substrate bulk properties deeper within
the slab. In [110]-direction four layers at the very bottom
of the slab are fixed in order to maintain bulk structure.
The atoms are labeled according to their species, thus each atom
can be defined to be either of $Au$ or $Pd$.
Rows in top layers are removed to create $(1 \times 2)$
and $(1 \times 3)$ missing row structures.

Before optimizing the surface structure, the slab is allowed to
expand uniformly to optimize the bulk free energy of $Pd$ at
each temperature separately.
After this the surface free energy is minimized by allowing local
relaxations using Polak-Ribiere algoritm.
The surface free energy is calculated by summing up
the free energies of the atoms in
the first eight layers subtracted by the bulk free energy of $Pd$
per atom and dividing the sum by the area of the surface.

The surface free energies of the following ordered structures
were studied: a bare $(1\times1)$ surface of $Pd$, $(1\times 2)$
reconstructed structure with $\Theta = 0.5ML$ of $Au$ and
an unreconstructed surface with $\Theta = 1ML$ of $Au$.
For $\Theta = 1.5ML$ an ordered
$(1\times2)$ structure with a concentration profile $(1,1,0)$ is
studied as well as an unreconstructed structure with a profile
$(1,{1 \over 2},0)$ with a disordered second layer
(and with non-zero configurational entropy). In the case of
$\Theta = 2ML$ there are four simple ordered structures :
$(1\times 1)$ with the profile $(1,1,0)$ \cite{Pro} (Fig. 1. (a)),
$(1\times 2)$ with the profile $(1,1,{1 \over 2})$ (Fig. 1. (b)) and
two possibilities for $(1\times 3)$,
either with the profile $(1,1,1)$ (Fig. 1. (d))
or $(1,1,{2\over 3},{1 \over 3})$ (Fig. 1. (c)) with
$Au$ distributed along the saw-tooth shaped
pattern forming effectively one atom thick layer along
the corrugation. The latter $(1\times 3)$
configuration proved more favourable. An unreconstructed layer
with $\Theta = 3ML$ of $Au$ is compared with
the corresponding structure of $\Theta = 2ML$ to see whether
layering on the top of the second layer should take place.
To discuss the concentration profile found in Ref. \cite{Kau}
for a $(1 \times 2)$ reconstructed film of $\Theta = 2.5ML$
two concentration profiles, $(1,1,1,0)$ and $(1,1,0,1)$,
are compared. In addition, some disordered structures for
$\Theta = 2ML$ are analyzed
to discuss the effect of configurational mixing entropy.

Five main observations to give a framework for the experimental
results can be found (See Fig. 2): {\bf (i)} At submonolayer
range $Pd$ atoms in the neighbourhood of the vacuum have a very
high energy. Thus the free energies of a bare $Pd$ surface
and a $(1 \times 2)$ surface with $\Theta = 0.5ML$ are quite
high in free energy compared
to $(1 \times 1)$ structure of a monolayer of $Au$.
Although the {\it bulk} free energies of $Au$ and $Pd$ are
quite close to each other, the relatively
long range of $Pd-Pd$ interactions make it energetically very
expensive to expose $Pd$ atoms to the vacuum. One monolayer
with $(1 \times 1)$ structure,
whose surface free energy is used as the reference
in Fig. 2, is the stablest structure up to
room temperature where $(1 \times 2)$ reconstructed surface with
$\Theta = 1.5ML$ becomes virtually equally stable.

{\bf (ii)} For $\Theta = 1.5ML$ at low temperatures a
reconstructed $(1\times 2)$ surface has a lower energy than
an unreconstructed surface (Fig. 2.) but,
the low energy structure $(1 \times 1)$ with $\Theta = 1ML$
may lead to a possibility of $(1 \times 1)$ domains
at low temperatures.
Increasing temperature beyond $T \approx 500K$ the reconstructed
and unreconstructed surfaces with $\Theta = 1.5ML$ have virtually
equal surface free energy. However, the unreconstructed structure
should gain in configurational entropy at high temperatures,
since the randomness of $Au$ atoms in the second layer
has little cost in free energy. Taking into
account this effect, it is to be expected that
a LEED pattern of $(1\times 2)$ structure for $\Theta = 1.5ML$
is observed at some narrow temperature range (as in Ref. \cite{Smi}).

{\bf (iii)} For $\Theta = 2ML$ there is a sequence of transitions
as seen in Fig. 2.:
$(1 \times 1) \rightarrow (1\times n) \rightarrow (1\times 2)
\rightarrow (1\times 3) \rightarrow  (1\times 1)$
with $n$ being a mixture of $1,2$ and $3$,
which is something very similar to what was observed in
Refs. \cite{Smi,Kau}, apart from incorrect temperatures. The
phase diagram is complicated due to
many factors contributing to
the thermal free energy, obviously having very different
temperatures of dominance. For example, the higher
order the reconstruction is, the more corrugated
the surface, and increasing corrugation makes
an $Au$ film effectively thinner, thus bringing
$Pd$ atoms closer to the vacuum, which increases surface
free energy. On the other hand, corrugation
decreases the proportion of $Au$ atoms at sites with low coordination,
which decreases surface free energy. In addition, missing rows
alter the local Einstein frequencies, which also has some
complicated effects on free energy. Some care must be taken
in interpreting these results, since the coverage is thick enough
to facilitate interdiffusion.

{\bf (iv)} Coverages beyond $\Theta =2ML$ are energetically
expensive except at high temperatures, where $\Theta =2.5ML$
is seen to decrease surface free energy (Fig. 2). Although not
shown in Fig. 2, the unreconstructed $\Theta =3ML$ surface is
very high in free energy throughout the temperature range
and the same applies to $(1 \times 3)$ reconstructed film
although the latter is clearly less expensive than
the unreconstructed surface at high temperatures. This result
suggests that layer-like growth beyond $\Theta =2ML$ does not
take place and the faint LEED patterns
at thick coverages \cite{Smi} are caused by clustering
of $Au$ on the top of the first two layers.

{\bf (v)} At low temperatures both concentration profiles,
$(1,1,1,0)$ and $(1,1,0,1)$ for $(1\times 2)$ reconstructed
film with $\Theta = 2.5ML$ are virtually equally expensive
in free energy (see Fig. 2). Both the
configurations approach lower coverages in surface energy
as the temperature is increased with a deviation in favour
of the profile {\it without} an intermediate $Pd$ layer. However, the
difference is so small that inclusion of intermixing could
change the figure. Thus the present analysis does not rule out
the possibility of intermediate layers with high concentration of
$Pd$ at high coverages as suggested in Ref. \cite{Kau}.
The main condition for such an intermediate layer to
exist is a thick $Au$ coverage that prevents $Pd$ atoms from being
exposed to the vacuum.

The main source of inaccuracy here is the absence of intermixing,
which means ignoring configurational mixing entropy.
For alloys, it is possible to deal with the mixing
entropy using Bragg-Williams type approximation,
as is done in Ref. \cite{Sutb}. In that study the ensemble is
grand canonical, while here it is canonical due to
a fixed coverage. But even in this case the mixing
entropy per layer per unit area for
randomly mixed layers can be approximated as follows:
\begin{equation}
S^{k} = -nk_{B}[c_{k} ln{(c_{k})} + (1-c_{k}) ln{(1-c_{k})}]
\end{equation}
where $c_{k}$ is the concentration of $Au$ atoms in the $k^{th}$
layer and $n$ is the number of atoms per unit area.
Disorder may increase vibrational free energy,
but the increase may be overcome by the contribution
of mixing entropy, $\sum_{k} TS^{k}.$

The effect of mixing entropy is roughly estimated
for $\Theta = 2ML$ by calculating the free energy,
Eq. (1) and the entropy, Eq. (3), for a couple of
energetically favourable concentration profiles with disorder.
For an unreconstructed surface
the profile $(1, {1\over 2}, {1\over 2})$ has a low cost
in free energy despite two randomly mixed layers
(The $Au$ atoms of the $2^{nd}$ $Au$ layer
are diluted to the $2^{nd}$ and the $3^{rd}$ layers).
For $(1 \times 2)$ reconstructed surfaces a corresponding
low cost/high entropy profile would be $(1,1,{1\over 2})$,
but with only one randomly distributed layer, leading to
a half of the entropy of $(1 \times 1)$ surface.
For $(1\times3)$ surface only the atoms in the $4^{th}$ layer of the
profile $(1,1,{2\over 3},{1 \over 3})$ are allowed to be
redistributed without a high expense, and the gain in entropy
is slightly smaller than for the mixed $(1 \times 2)$ structure.
In Fig. 3 are shown the lowest limits of free energies corrected
by mixing entropies for mixed layers with
for randomly distributed $Au$ atoms as described above.
The complicated transition sequence survives the
error analysis, because the configurational entropy
affects to the same direction for all the structures, although
favouring $(1 \times 1)$ surface and disfavouring $(1 \times 3)$
surface. This might change the sequence of transitions more like
$(1 \times 1) \rightarrow (1\times n) \rightarrow (1\times 2)
\rightarrow (1\times 1),$ which actually is what Ref. \cite{Smi}
suggests. In any case, the effect of configurational
entropy is significant above room temperature.

Keeping in mind the restrictions of the model the present study
gives results in qualitative agreement with
experimental observations. The calculations do not merely reproduce
the structural phase transitions observed by LEED but also
provide explanations to some of the experiments. The calculations
take only a couple of minutes per point
on an HP workstation and thus the method can easily be used
in planning experiments and interpreting LEED or STM results.
Including intermixing and optimization of the concentration
profile to the model is a logical and necessary continuation
to the present work.

\par

{\bf Acknowledgements: }
I would like to thank Dr. A.P. Sutton, Dr. M. Lindroos, Dr. C. Barnes
and Mr. P. Kaukasoina for helpful discussions.
The author has been funded by the Academy of
Finland and the Royal Society. The inspiring atmosphere
of Materials Modelling Laboratory (partially funded by
SERC grant No. GR/H58278) of the Department of Materials in Oxford
University is gratefully acknowledged.

\begin{figure}
\caption{Different surface structures for $\Theta = 2ML$ of
$Au$ (black circles) on Pd(110) surface (open circles):
(a) $(1\times 1)$, (b) $(1 \times 2)$, and (c) and (d)
$(1 \times 3)$ structures as seen from the side of the
slab in the direction of rows. The structure of (c) has
lower free energy than the structure of (d).}
\end{figure}

\begin{figure}
\caption{Excess free energies for different coverages and surface
structures, the surface free energy of $(1 \times 1)$ structure of
$\Theta = 1ML$ being the reference. For $(1 \times 3)$, the structure
of Fig. 1 (c) has been used. In the legend, $2.5ML$ (non.) refers
to the nonmonotonous concentration profile of $\Theta = 2.5ML$.}
\end{figure}

\begin{figure}
\caption{The error analysis for excess free energies of different
structures for $\Theta = 2ML.$ The symbols without guiding
lines show the lower limits of the free energy as correted with
the configurational entropy.}
\end{figure}

\end{document}